# Support of Study on Engineering Technology from Physics and Mathematics

**Djafar K. Mynbaev**

*Electrical Engineering Technology and Telecommunications Department, New York City College of Technology, CUNY, Brooklyn, NY 11201*

**Candido Cabo,**

*Computer Systems Technology Department, New York City College of Technology, CUNY, Brooklyn, NY 11201, USA*

**Roman Ya. Kezerashvili,**

*Physics Department, New York City College of Technology, CUNY, Brooklyn, NY 11201, USA*

**Janet Liou-Mark**

*Mathematics Department, New York City College of Technology, CUNY, Brooklyn, NY 11201, USA*

## Abstract

An approach that provides students with an ability to transfer learning in physics and mathematics to the engineering-technology courses through e-teaching and e-learning process is proposed. E-modules of courses in mathematics, physics, computer systems technology, and electrical and telecommunications engineering technology have been developed. These modules being used in the Blackboard and Web-based communications systems create a virtual interdisciplinary learning community, which helps the students to transfer knowledge from physics and mathematics to their study in engineering technology.

## 1. Introduction

Engineering is the science that made the properties of matter and the sources of power useful to humans. Technology turns engineering ideas into practical devices, structures, machines, and products. People working in engineering technology apply the principles of physics and mathematics, among others, to practical technical problems. Since engineering relies on the application of mathematics and science to development of useful products or technologies, it is engineering that turns ideas into reality. At the same time, the practical needs met by engineering technology stimulate further development of physics and mathematics. Therefore, engineering technology, physics and mathematics are interconnected and work in unison.

Mathematics, being one of the foundations of engineering, has, on the other hand, always been one of the largest stumbling blocks, causing first-year students to drop out of engineering programs[1]. Even most engineering-technology students understand that having a good foundation in mathematics is essential to successfully complete a degree in their program, they struggle with the discipline and often give up pursuing the major.

Physics is the study of the physical world and physics is an indispensable component in engineering curricula because technology is based on our knowledge of physical laws. Physics remains the leader of the modern natural sciences, the theoretical basis of modern engineering and, as no any other science, promotes the development of creative and critical thinking in future engineers. Good training in physics also provides a solid foundation for lifelong learning. However, research in physics education in different countries shows that students at the college and even university levels continue to hold fundamental misunderstandings of the world around them. Science learning remains within the classroom context and just a small percentage of students are able to use the knowledge gained at school for solving various problems of the larger physical world [2,3]. In most of the courses, students listen to lectures without strong connections to their everyday experiences. Students usually do not have the opportunity to form their own ideas; they rarely get a chance to work in a way where they are engaged in discovery and building and testing models to explain the world around them, like the scientists do.

One of the problems in engineering technology education is that the students do not correlate closely engineering and science subjects and do not always transfer the knowledge they acquired in physics and mathematics to their engineering-technology classes. This problem is addressed in our project by developing a close relationship among faculty teaching engineering-technology courses and those teaching physics and mathematics. The goal of our approach is to provide students with the ability to transfer readily the material studied in physics and mathematics to the engineering-technology courses. Simultaneously, an engineering-technology student studying physics and mathematics will have access to applications in the engineering-technology field.

A two-step approach has been developed to solve this problem: First, the student has access to interdisciplinary materials in engineering, mathematics, and physics. For example, a student studying a course in electrical-engineering technology will have access to course modules that cover the proper topics in physics and mathematics. Moreover, a student can use a virtual physics laboratory to perform specific experiments related to a theory. Secondly, by creating virtual space, that is, a place on the Web, a student can easily navigate from one subject to another. To achieve this goal, e-modules of courses have been developed with cross-references between common key terms and definitions. If a student in electrical-engineering technology needs to review a specific concept or topic in physics or mathematics, the e-modules will provide the appropriate linkage. To support this service, a Web site accessible to the students has been constructed.

This paper discusses, in case-study format this approach, its achievements, and problems encountered in developing this project.

## 2. The Need for More and Better Engineers

The health and prosperity of the American economy relies on high productivity of its workers. The key to productivity is innovation, particularly the ability to develop new technology. Clearly, the more brainpower we have, the more ideas—that is, the more innovations—we will get. Thus, there is a strong need to increase the number of engineers that our country turns out. And the United States, a world leader in higher education, doesn't lead the world in terms of the number of engineers produced by its universities. According to Internet reports, China educates about 600,000 engineers annually, while India graduates about 350,000 engineers each year. On the other hand, American colleges award only some 69,000 bachelor's degrees in engineering annually. These numbers are shown in Figure 1 with the indication of the percentage of engineering graduates in these countries with respect to their population. Of course, other important factors, such as quality of education and the level of country development, must be taken into account for the comprehensive analysis of the problem, but these numbers show the relative volume of brain power produced by two main fast-developing countries. We need to bear in mind that other developed and developing countries also educate a great number of engineers to fully satisfy their world. In contrast, according to Mark Hurd, chairman and CEO of Hewlett-Packard, the world's largest technology vendor in terms of sales, the United States "is graduating more sports management professionals than engineers, which isn't good in a global economy where innovation can make or break a company[4]."

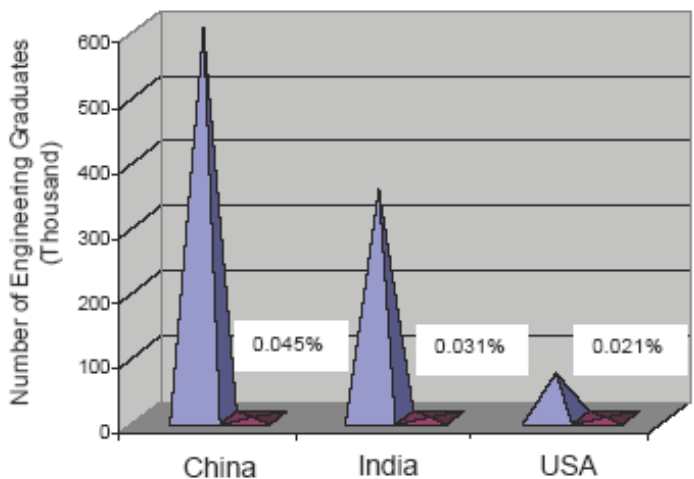

Fig. 1. Number of engineering graduates each year and their percentage in China, India and USA.

The example of solution of this problem is the situation in California. That state currently faces a significant shortage of engineers. What's more, the state's Labor and Workforce development Agency (LWDA) projects that the state will have a shortfall of almost 40,000 engineers by 2014. There will be a need for approximately 20,000 to 24,000 additional engineers in California to fill the gap between demand and supply of engineering educates in both the private and public sectors over the next decade. Thus, California Governor Arnold Schwarzenegger announced an initiative to educate 20,000 new engineers in his state. This initiative includes expanding the number of high schools that offer extensive engineering-focused programs designed to better prepare the students for entering engineering programs in universities. In addition, he wants to establish special program at the University of California and at the California State University system to certify military veterans with an engineering background and to attract more private funds to develop partnership between California community colleges and private industry; and some other steps[5]. Obviously, not only the number but also the quality of engineers is a main concern. Many experts say that we need "to provide funding and freedom for local universities to create the best engineering schools possible…."[6] Industry and government leaders fully understand this issue and have started to take practical steps to resolve it. At the national level, the federal government recently commissioned a group of industry leaders to analyze the status of engineering education in our country and suggest steps to improve the situation.

These examples show that the need to improve engineering education in both quantitative and qualitative aspects at every level—national, state and local—is well understood; what the problem boils down to is practical steps to achieve the well-defined objective. This is why federal agencies support all attempts in this regard and the work at New York City College of Technology (City Tech) is based on such support.

## 3. Student Population at City Tech and Our Task

It's well known, of course, that in the United States there are engineering and engineering-technology schools. Graduates of engineering schools traditionally work as researchers, developers, and designers of new devices and technologies. Graduates of engineering-technology schools work primarily as maintenance and control personnel; that is, they operate existing equipment rather than create new equipment. Though the trend is gradually eliminating the fine line between design and maintenance, the traditional separation between these two types of work still exists.

City Tech, where this work is under way, is an urban college that provides engineering-technology education and student population is predominantly minority and many are the first members of their family to pursue a college degree. In the fall of 2007, for instance, City Tech served 13,506 students, more than 65% of whom were minority. Black (non Hispanic) students comprised 36.8% of the student population; Hispanic students comprised 28.6%; Asian/Pacific Islanders comprised 16.3%; and Caucasians (non Hispanic) comprised 11.3%. Interestingly, 60.6% of our students report language other than English spoken at home.

## 4. How to Teach Physics and Mathematics to Engineering-Technology Students?

The study of engineering technology relies on material that students learned in their physics and mathematics courses. The problem is that the majority of our students do not correlate closely engineering and science subjects and do not always transfer the knowledge they acquired in physics and mathematics to their engineering-technology classes. As a result, the students get lost in engineering-technology courses when the topic under study requires reference to specific knowledge and skills taught in a mathematics or physics courses. As a result, the engineering-technology professors meet the situation in which the students formally studied a course but did not interrelated learning material with engineering applications.

This problem raises a valid question: How do you teach physics and mathematics in such an environment? One approach is to incorporate knowledge delivered in physics and mathematics courses into engineering-technology courses to integrate these knowledge and skills with their applications. This approach implies that one instructor teaches all subjects, a pursuit that bears obvious shortcomings. The main drawback of such an approach is that the student studies not physics and mathematics but the application of a set of rules and skills to a specific technological field. In the traditional approach, physics and mathematics are taught by experts in these disciplines. This, however, very often creates the problem we encounter in our technology courses. Nonetheless, physics and mathematics professors deliver the "bread and butter" of these subjects thereby meeting the educational goals of any academic institution.

One approach that aims to resolve this dilemma has been taken by Franklin W. Olin College of Engineering[7], where the curriculum, on the one hand, offers separate courses in mathematics and physics and, on the other hand, increases the emphasis on interdisciplinary design, which implies "learning through doing." To understand whether we can transfer this approach to other institutions, we have to bear in mind that Olin can afford such an innovative curriculum because it is a highly selective academic institution; for example, class 2008 lists 68 students admitted out of 546 applicants and it has a 9:1 student-to-faculty ratio. Olin recently graduated its first clan of engineers and the effectiveness of its approach remains to be tested, depending of course on the success of its graduates.

City Tech adheres to the traditional model of teaching physics, mathematics and engineering technology. However, in our, career-oriented college, physics and mathematics must be closely connected to and support their applications in engineering-technology courses. This is the problem we are trying to solve.

## 5. Our Approach

We think that the main problem is not a lack of foundational knowledge, but the inability of students to transfer that knowledge from mathematics and science to the technologies. Many students think of their college career as a number of disconnected courses that need to be passed in order to graduate. As explained earlier, our approach to facilitate knowledge transfer is to use the paradigm of the Web and its syntax and hypertext to create networks of linked documents that allow students to move from educational material in the technologies to educational material in mathematics and physics. The architecture of the system is as follows[5]: We have two systems—open and closed. A Web-based system accessible to all instructors and students is open one. The Blackboard accessible only to instructors who teach these courses and to students who register for these courses is closed system. We place our courses on both systems. Each course on the Web site in physics, mathematics, computer-system technology (CST), and electrical and telecommunication engineering technology (EET) is linked to a corresponding course on Blackboard. By logging into Blackboard or Web site, the student becomes a participant in the virtual classroom. This architecture is shown in Figure 2; a detailed discussion of our system is presented in a paper[8].

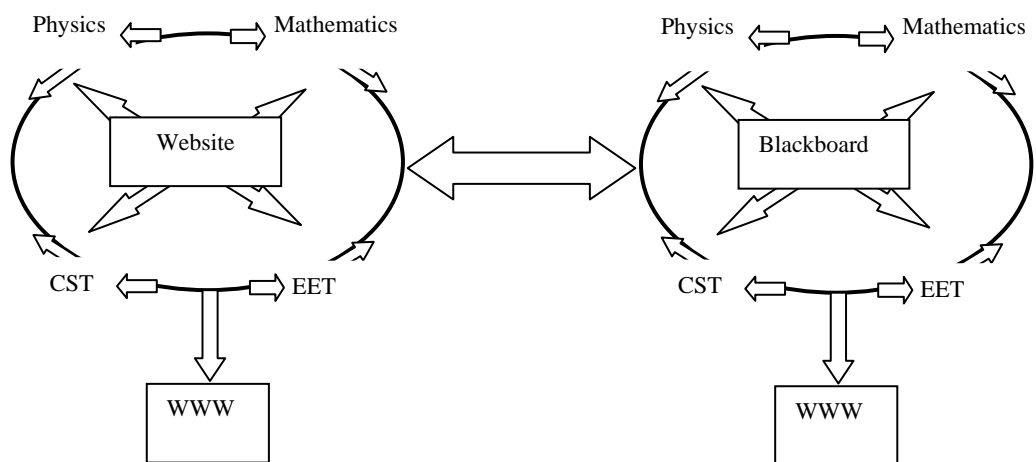

Fig. 2. Architecture of the e-learning and e-teaching system[8].

The courses placed on Web site and Blackboard has cross-references. For instance, a student studying the application of Ohm's law in a course on electrical circuit needs to solve a linear equation. When words "Ohm's law" or "linear equation" appear in this course, a student, by clicking on these words, will get the link to the corresponding lecture material in physics or mathematics where a detailed explanation of these concepts and methods to solve these equations will be found. On the other hand, by studying, let's say, "Total Internal Reflection" in a physics course, the student can link to an application of this concept in fiber-optic communications and therefore immediately realize that this concept has real application to the student's major.

The Website and Blackboard are also linked to WWW. Instructors establish these links relate to course material to the best available and most suitable pertinent information on the Web. At the same time, by logging into Blackboard, the student becomes a participant in the virtual classroom. Here the student finds all the material necessary for successfully completing the course. Thus, this approach utilizes a course management tool, Blackboard, which provides innovative ways to teach, learn, communicate, and collaborate across disciplines. By putting courses online, instructors of different disciplines now strive to create a networked learning environment among different departments.

## 6. Case studies

### Computer Systems Technology

As part of the requirements of the bachelor of technology offered by the Department of Computer Systems Technology at City Tech, students have to complete three sub specialization modules in different areas of computer systems. Each module is a sequence of three courses in areas such as computer programming, systems analysis and design, database design, networking, Web programming and design, and computer security. Our program is different from a traditional computer science program because our emphasis is on the application of emerging information technologies to business, science, the arts and other human endeavors, and not only on the theory of computation. Still computer technologists need an in-depth understanding of theory to be able to effectively design and implement applications. Our instructional approach could be described as "hands-on + theory", where the theory guides the development of applications, and application development reinforces the understanding of the theory.

The theoretical aspects of computer technology rely on fundamental concepts of mathematics and the physical sciences. Therefore, instructors in computer courses assume that students are familiar with those fundamental mathematical and physics concepts. However, often, this is not the case, and instructors have to spend time reviewing concepts that students should have already mastered.

It is possible that some students may not have the basic foundations in mathematics and physics required for computer courses because they never study the material. About half of our junior-level students are transfer students who completed their associate degree elsewhere; the junior year is their entry point into our program. While we maintain articulation agreements and frequent conversations with our main feeder programs at other colleges, students come with a heterogeneous background in mathematics and science. Some students come to our specialization modules having had pre-calculus, others having had a course in statistics, and others having had discrete mathematics. The background in science is still more diverse: some students had taken physics; others, biology or chemistry. Some had not taken any science course at all.

In the area of computer systems, so far we have created online educational material for the networking module (one of our specialization modules), which consists of three courses: CST 3507 (Advanced Single Local Area Networks Concepts), CST 3607 (Interconnectivity), and CST 4707 (Local Area Network – Internet Connection). The next step is to identify concepts (keywords) that require background knowledge in mathematics and physics, and link those keywords to the corresponding mathematics and physics educational material. For example, IP addressing is an important concept that is discussed in CST 3607; it requires an understanding of binary arithmetic. A link in the educational material on IP addressing will be created to the educational material explaining binary arithmetic in mathematics. Of course, the expectation is that by using hypertext, we will encourage our students to read educational material not only in a traditional (textbook) linear fashion but also to explore lateral connections, which include background material or other applications. It is our expectation that the lateral connections will facilitate and reinforce the students' understanding of the primary material.

The same strategy proposed here to transfer knowledge between mathematics and physics and computer systems can be used to transfer knowledge between different areas of computer systems. Students majoring in computer systems technology often have problems transferring knowledge and skills from one specialization module to the other or between courses with the modules. (See above for the modular organization of the curriculum in our program.) For example, the sub-specialization module in Web design and programming makes use of concepts in programming, databases, and networking. Many students have trouble applying the concepts they have already learned in the database and/or networking module to a similar problem in a different context of the Web. We believe that using the approach to interdisciplinary knowledge transfer which is described will enable us to enhance such a transfer within our major.

## Electrical and Telecommunications Engineering Technology

The Department of Electrical and Telecommunications Engineering Technology offers associate-degree programs in electrical engineering technology and telecommunications engineering technology and a bachelor's-degree program in telecommunications engineering technology. To educate our students properly, we are working to develop more effective methods to transfer knowledge students learned in their physics and mathematics courses to the same students when they enroll in our technical programs. To implement this goal, we participate in the interdisciplinary work described in this paper. Specifically, we have developed e-modules of our courses and we are in the process of identifying key words to establish links with courses in physics and mathematics.

The specifics of our work in this project can be described as follows[8]: Electrical engineering technology modules comprise, at this stage, three courses: Network Analysis I, Network Analysis II and Electronics. These courses include analysis of dc and ac circuits and basic electronics; they are, in essence, the first courses that electrical engineering technology students meet in their major. The courses Network Analysis I and II combine both theory and laboratory segments; the Electronics course is a theory course only; laboratory exercises for this course are conducted in a separate class.

Engineering-technology programs always rely on hands-on experiments and our circuit-analysis courses can serve as good examples of such an approach. During one week of study, the students learn a specific topic in theory and confirm this concept via experiment. For example, students study Ohm's law during a traditional lecture and then confirm its validity by measuring a voltage-current relationship. In our relevant module, we provide students with a derivation of the basic formula and discuss this formula in various formats. The focus of this study is the application side of Ohm's law; let say, in one format we concentrate on how current flowing through a resistor changes if the applied voltage varies. In a corresponding experiment, students make actual measurements of current while changing voltage.

With this study, students grasp the concept of Ohm's law. To introduce this law (and other topics, for that matter) from a different standpoint, we also use computer-based experiments.
Electrical engineering technology modules include, of course, sets of problems and solutions. In our example of Ohm's law, again, students practice computing values of current while varying applied voltage with a given resistance. They also learn through a problem-solution approach how the value of a resistor changes the slope of a voltage-current graph. This approach significantly enhances the students' learning ability.

Developing of e-modules allows building links with courses in mathematics and physics through key words highlighted in courses. Continuing our consideration of Ohm's law as an example, we can refer to the word "resistance," which students inevitably encounter in studying this topic. By clicking on this key word, the students will easily transfer to the physics course where this phenomenon is discussed in details. At the same token, when considering the formula for Ohm's law, our students will see the words "linear equation." By clicking on these words, the students will switch to the proper e-module of mathematics, where this topic is considered from the mathematical standpoint. The important point here is that the students will see not general information about resistance and linear equation but the specific discussion that they already had during their course of study in these physics and mathematics courses. In other words, when the students switch to e-modules in different disciplines, they will find themselves in "familiar territory"; they will be able to review the material they have already studied. We strongly believe that such a "repetition" will greatly improve their ability to cope with their technical courses.

Certainly it is important to expand this approach to establish links among all disciplines that students study within their major. When completed, this work will help to provide a close relationship not only among different disciplines but also within every major.

### Mathematics

Students entering the engineering-technology program at City Tech are required to take and pass Mathematical Analysis (MAT 1375), a pre-calculus functions course. This requirement produces more pressure on students to be pre-calculus-ready when they enter the college, which many are not.

Because many of City Tech students are not well prepared to enter the field of engineering technology, e-modules for two bottleneck courses in mathematics were designed. The first course is MAT 1275 (Introduction to Mathematical Analysis) is the prerequisite course for pre-calculus. The second course, naturally, is MAT 1375 (Mathematical Analysis). MAT 1275 covers topics ranging from intermediate to advanced algebra, including quadratic equations, systems of linear equations, exponential and logarithmic functions. In addition, this course includes such topics from trigonometry as identities, equations and the solution of triangles. MAT 1375 includes topics from advanced algebra and theory of equations, such as the solution of polynomial equations, DeMoivre's Theorem, binomial theorem, vectors, lines, conic sections and progressions.

At City Tech, students are most often overwhelmed by the materials covered in each mathematics class. Because of the heavy content load and poor background skills, their main focus is to understand the mathematical concepts. Whether they connect these concepts with other subjects areas will depend on the instructor's examples. Instructors are challenged because they have to work with some forty students with mixed abilities and from diverse disciplines. It is virtually impossible to focus mathematics on a particular area, such as engineering. This may be one of the main reasons why students have difficulty transferring their knowledge of mathematics to engineering, computing systems, and/or physics. To resolve this ongoing problem, e-modules interconnecting mathematics, engineering, physics, and computer systems are designed.

The design of mathematics e-modules is based on Bloom's taxonomy of cognitive educational objectives[9]. The purpose of this classification system is to categorize the cognitive changes produced in the students as a result of the methods being used. These cognitive objectives, from the simplest behavior to the most complex, are as follows: knowledge, comprehension, application, analysis, synthesis, and evaluation.

Since one of the main goals of the project is the transfer of knowledge of mathematics to engineering technology, the e-modules have been created to support each cognitive objective. After perusing each e-module, a student should know how to define the vocabulary associated with the mathematics concept (knowledge), master the computational skills by going through the examples (comprehension), solve real-word problems associated with the topic (application), explain the mathematical process in deriving the answer (analysis), apply prior knowledge and skills to produce a new concept (synthesis), and compare mathematical methods (evaluation).

The versatility and availability of the e-modules allow engineering-technology students to connect mathematics to the courses required in engineering. Each e-module can be accessed by students at any time and anywhere as long as they have Internet access. The modules cover the mathematics theory in a form that is easy to understand and they include examples worked out in other disciplines. The lessons can be studied independently, allowing students to work at their own pace. Consequently, they are ideal for reviewing and self-learning.

### Physics

Physics cannot be taught only using the book and blackboard and asking students to memorize rules, formulas and laws. One of the important parts of teaching physics is a real-time experimental demonstration that visualizes the laws of nature; laboratory exercises that the students conduct during laboratory sessions serve the same purpose. The laboratory and experimental demonstration should be established as a primary learning tool in the science, technology, engineering and mathematics at an early point in students' academic careers so that students have a taste of the excitement of science and engineering research. Indeed, one of the places where active and collaborative learning and applications of the laws of nature can be realized is the physics laboratory, where students become active participants of the learning process [10-13].

Therefore, it is necessary to promote teaching and learning resources which supports "Instructor-Student-Experiment" interactive engagement and that does a physics module of our system.

Physics e-modules of the system, which contain two courses of algebra-based general physics (PHYS 1433 and PHYS 1434), reflected all of the above mentioned features and included several teaching resources aimed to promote comprehension of the physics laws: class lecture, demonstration experiments, laboratory experiments, e-learning material, problem-solving sessions. Class lecture presents one of the most important principles for every physics course, - *concepts first*. Conceptual understanding is the focus through the explanations, examples and media demonstrations of the experiments and is presented on the Website and on the Blackboard. E-learning material provides problem solving examples, and problem-solving session provides to students through the interactive system between student and instructor **"Physics Tools"** [14], which is a Problem-Based Learning and Problem-Solving tool, given on the website. The architecture of our system is implemented primary in HTML with embedded Java scripts and Java applets. The system allows students to check the analytical solution of the problem, numerical value and units, provides the hints and the detail solution of the problem by a request, evaluates a student's work and gives the correct answer by a request. Physics Tools module consists from two shells. One shell is intended for students and other is intended for an instructor. Physics Tools is designed to help students to understand main principles and concepts of physics through visualization experimental phenomena and performing virtual experiments and developing problem-solving skills. The objectives of Physics Tools are to improve the success rate of students enrolled in algebra and calculus-based physics classes and to develop critical thinking skills. This system provides students an outstanding training and preparation basis - exactly what they need in order to score well on typical physics exams.

The physics modules provide demonstrations of real physics experiments from mechanics, liquids, oscillations and waves, electricity, magnetism and optics. Each experiment is a stand-alone unit, which illustrates a particular principle of physics; it is also closely related to the lecture materials. In developing this module we are using existing 25 DVDs of the video encyclopedia of physics demonstrations [15], some demonstration suggested by us [16,17] and digital taping them as well as making the links to existing Physics Lecture Demonstrations on Internet [18]. Most video experiments have parallel soundtracks and detail descriptions. All these will permit students to listen and read the explanation of the physics phenomena. Student may repeat the experiment few times and concentrate on learning and understanding the physical processes. All these allow for interactive participation by students for studying and understanding physics concepts and principles and their applications in engineering. We have also revised physics laboratory curricula and developed physics laboratory exercises that support engineering applications.

## 7. Conclusion

During the first year of this project, almost three thousand City Tech students benefited from the project by virtue of fact that their science, technology, engineering, and mathematics professors participated in faculty-development activities. As a result, their students had the opportunity to learn using laboratories and curricula materials developed over this year. As the virtual community grows in scope, the student population that will benefit from this project, especially underrepresented minorities and women, will increase, we hope, exponentially. To achieve this goal we are planning to finish building our Web site with complete cross-references among interdisciplinary courses and courses within every major.

We believe that our approach in building a close cooperation among different but closely related disciplines by creating a virtual learning community will result in greater academic success of our students.

*Acknowledgement:* This work is supported by US Department of Education grant P120A060052.